\documentclass[12pt]{amsart}
\usepackage[all]{xy}
\newif\ifmypdf
\mypdffalse
\usepackage[\ifmypdf pdftex\else dvips\fi]{graphicx}

\newcommand{\ed}{

\end{document}
}

\newcommand{\Includegraphics}[2]{
\ifmypdf
\includegraphics[#1]{#2.pdf}
\else
\includegraphics[#1]{#2.eps}
\fi}

      \newenvironment{changemargin}[2]{\begin{list}{}{
         \setlength{\topsep}{0pt}\setlength{\leftmargin}{0pt}
         \setlength{\rightmargin}{0pt}
         \setlength{\listparindent}{\parindent}
         \setlength{\itemindent}{\parindent}
         \setlength{\parsep}{0pt plus 1pt}
         \addtolength{\leftmargin}{#1}\addtolength{\rightmargin}{#2}
         }\item }{\end{list}}

\newcommand{\XYmatrix}[1]{\xymatrix@R=15pt@C=10pt{#1}}

\newcommand{\Nset}{{\{0,\dots,N-1\}}}
\newcommand{\ub}[1]{\underbrace{#1}}

\newcommand{\dom}{\op{dom}}
\newcommand{\la}{\langle}
\newcommand{\ra}{\rangle}

\newcommand{\be}{\begin{enumerate}}
\newcommand{\itm}{\item}
\newcommand{\ee}{\end{enumerate}}
\newcommand{\bi}{\begin{itemize}}
\newcommand{\ei}{\end{itemize}}
\newcommand{\sm}{\setminus}
\newcommand{\op}[1]{\operatorname{#1}}

\newcommand{\inv}{^{-1}}
\long\def\forget#1\forgotten{}

\newtheorem{thm}{Theorem}

\newtheorem{prob}[thm]{Problem}

\newtheorem{cor}[thm]{Corollary}

\theoremstyle{definition}
\newtheorem{defn}[thm]{Definition}
\newtheorem{exam}[thm]{Example}
\newtheorem{code}[thm]{Coding}

\theoremstyle{remark}
\newtheorem{rem}[thm]{Remark}

\title[Decomposing and iterating lookup tables]
{Decompositions of graphs of functions and fast iterations of lookup tables}
\author{Boaz Tsaban}
\thanks{Supported by the Koshland Center for Basic Research}
\address{Department of Mathematics,
Weizmann Institute of Science, Rehovot 76100, Israel}
\email{boaz.tsaban@weizmann.ac.il}
\urladdr{http://www.cs.biu.ac.il/\~{}tsaban} \keywords{fast
forward functions, fast forward permutations, cycle decomposition}

\begin{document}
\begin{abstract}
We show that every function $f$ implemented as a lookup table can
be implemented such that the computational complexity of evaluating $f^m(x)$ is small,
independently of $m$ and $x$.
The implementation only increases the storage space by a small \emph{constant} factor.
\end{abstract}

\maketitle

\section{Introduction and Motivation}

According to Naor and Reingold \cite{NaRe},
a function $f:\Nset\to\Nset$ is
\emph{fast forward} if for each natural number $m$ which is polynomial in $N$,
and each $x=0,\dots,N-1$, the computational complexity of evaluating $f^m(x)$---the
$m$th iterate of $f$ at $x$---is
small (polynomial in $\log N$).
This is useful in simulations and cryptographic applications,
and for the study of dynamic-theoretic properties of the function $f$.

Originally this notion was studied in the context of pseudorandomness,
where $N$ is very large -- see \cite{NaRe, ffperms, GGN}.
Here we consider the remainder of the scale, where $N$ is not too large,
so that the function $f:\Nset\to\Nset$ is or can be implemented
by a lookup table of size $N$. Implementations as lookup tables are standard for several
reasons, e.g., in the case where the evaluation $f(x)$ is required to be efficient,
or in the case that $f$ is a random function, so that $f$ has no shorter definition
than just specifying its values for all possible inputs.
We describe a simple way to implement a given function $f$
such that it becomes fast forward.
The implementation only increases the storage space by a small constant factor.

\medskip

The case that $f$ is a permutation is of special importance and is easier to treat.
This is done in Section \ref{SecPerm}.
In Section \ref{SecFunc} we treat the general case.

\section{Making a permutation fast forward}\label{SecPerm}

We recall two definitions from \cite{ffperms}.

\begin{defn}\label{OCS}
Assume that $f$ is a permutation on $\Nset$.
The \emph{ordered cycle decomposition} of $f$ is the sequence
$(C_0,\dots,C_{\ell-1})$ consisting of all (distinct) cycles of $f$, such that
for each $i,j\in\{0,\dots,\ell-1\}$ with $i<j$, $\min C_i < \min C_j$.
The \emph{ordered cycle structure} of $f$
is the sequence $(|C_0|,\dots,|C_{\ell-1}|)$.
\end{defn}

The ordered cycle decomposition of $f$ can be computed in time $N$:
Find $C_0$, the cycle of $0$. Then find $C_1$, the cycle of the first element not in $C_0$,
etc. In particular, the ordered cycle \emph{structure} of $f$ can be computed in time $N$.

\begin{defn}\label{ffperm}
Assume that $(m_0,m_1,\dots,m_{\ell-1})$ is
the ordered cycle structure of a permutation $f$ on $\Nset$.
For each $i = 0,\dots,\ell-1$, let $s_i = m_0+\dots+m_i$.
The \emph{fast forward permutation coded by $(m_0,m_1,\dots,m_{\ell-1})$}
is the permutation $\pi$ on $\Nset$
such that for each $x\in\Nset$,
$$\pi(x) = s_i + (x-s_i+1 \bmod m_{i+1})\quad\mbox{where }s_i\le x< s_{i+1}.$$
In other words, $\pi$ is the permutation whose ordered cycle decomposition is
$$\pi=(\ub{0\dots s_0-1}_{m_0})(\ub{s_0\dots s_1-1}_{m_1})(\ub{s_1\dots s_2-1}_{m_2})\cdots
(\ub{s_{\ell-2}\dots N-1}_{m_{\ell-1}}).$$
\end{defn}

The assignment $x\mapsto i(x)$ such that $s_{i(x)}\le x< s_{i(x)+1}$
can be implemented (in time $N$) as a lookup table of size $N$.
As
$$\pi^m(x) = s_{i(x)} + (x-s_{i(x)}+m \bmod (s_{i(x)+1}-s_{i(x)})),$$
$\pi$ is fast forward.

\begin{code}\label{Pcode}
To code a given permutation $f$ on $\Nset$ as a fast forward permutation,
do the following.
\be
\item Compute the ordered cycle decomposition of $f$:
$$f=(\ub{b_0\dots b_{s_0-1}}_{m_0})(\ub{b_{s_0}\dots b_{s_1-1}}_{m_1})(\ub{b_{s_1}\dots b_{s_2-1}}_{m_2})\cdots
(\ub{b_{s_{\ell-2}}\dots b_{N-1}}_{m_{\ell-1}}).$$
\item Define a permutation $\sigma$ on $\Nset$ by $\sigma(x) = b_x$ for each $x=0,\dots,N-1$.
\item Store in memory the following tables: $\sigma$, $\sigma\inv$, the list $s_0,\dots,s_{\ell-1}$
(where $s_k=m_0+\dots+m_k$ for each $k$), and the assignment $x\mapsto i(x)$.
\ee
\end{code}
Let $\pi$ be the fast forward permutation coded by $(m_0,m_1,\dots,m_{\ell-1})$. Then
$$f = \sigma\circ\pi\circ\sigma\inv.$$
For each $m$ and $x$, $f^m(x)$ is equal to
$\sigma(\pi^m(\sigma\inv(x)))$, which is computed by $5$ invocations of the stored lookup tables and $5$
elementary arithmetic operations (addition, subtraction, or modular reduction).
We therefore have the following.

\begin{thm}
Every permutation $f$ on $\Nset$ can be coded by $4$ lookup tables of size $N$ each, such that
each evaluation $f^m(x)$ can be carried using $5$ invocations of lookup tables and $5$
elementary arithmetic operations, independently of the size of $m$.\hfill\qed
\end{thm}

\begin{rem}
~\be
\item For random permutations, $\ell\approx\log N$ and therefore the total amount of
memory is about $3N+\log N$.
\item Instead of storing the assignment $x\mapsto i(x)$, we can compute it online.
This is a search in an ordered list and takes $\log_2(\ell)$ in the worst case.
For a typical permutation this is about $\log_2(\log(N))$ additional operations in the worst case
(e.g., for $N=2^{32}$, this is about $4$ additional operations per evaluation).
This reduces the memory to $2N+\log N$.
\ee
\end{rem}

\section{Making an arbitrary function fast forward}\label{SecFunc}

We begin with a simple method, and then describe a twist of this method
which gives better results.\footnote{See new footnote \ref{newfootnote} on page \pageref{newfootnote}.}

\subsection{The basic approach}\label{SecFuncBas}
The language of graphs will be convenient.
For shortness, a (partial) function $f:\Nset\to\Nset$ will be called
a \emph{(partial) function on $\Nset$}.

\begin{defn}
Let $f$ be a partial function on $\Nset$.
The \emph{graph of $f$} is the directed graph $G=\la V,E\ra$,
where
\begin{eqnarray*}
V & = & \Nset,\\
E & = & \{(x,f(x)) : x\in\dom(f)\}.
\end{eqnarray*}
The \emph{orbit} of an element $v\in V$ is
the \emph{maximal simple} tour $(v,v_1,v_2,\dots,v_k)$ in $G$.
Note that either $f(v_k)$ is undefined, or else $f(v_k)\in\{v,v_1,v_2,\dots,v_k\}$.
In the latter case, we say that the orbit is a \emph{$\rho$-orbit}.
\end{defn}

Any subgraph of a partial function $f$ on $\Nset$
is the graph of some restriction of $f$, and in particular is
the graph of some partial function $g$ on $\Nset$.

\begin{defn}\label{OCSf}
Assume that $f$ is a function on $\Nset$.
The \emph{ordered orbit decomposition} of $f$ is the sequence
$(C_0,\dots,C_{\ell-1})$ defined by:
\be
\itm $C_0$ is the orbit of $0$.
\itm For $k>0$, if $V\neq C_0\cup C_1\cup \dots \cup C_{k-1}$,
then $C_k$ is the orbit of the least element of
$V\sm (C_0\cup\dots \cup C_{k-1})$ in the subgraph
induced by $G$ on the vertices in $V\sm (C_0\cup\dots \cup C_{k-1})$.
\itm $\ell$ is the least $k$ such that $V=C_0\cup\dots \cup C_{k-1}$.
\ee
The \emph{ordered orbit structure} of $f$
is the sequence $(|C_0|,\dots,|C_{\ell-1}|)$.
\end{defn}

Note that the ordered orbit decomposition of a permutation is just its ordered cycle decomposition.
Assume that $(C_0,\dots,C_{\ell-1})$ is the ordered orbit decomposition of $f$.
Clearly, $(C_0,\dots,C_{\ell-1})$ can be reconstructed from
the concatenated sequence $C_0C_1\cdots C_{\ell-1}$ together with the
ordered orbit structure $(|C_0|,\dots,|C_{\ell-1}|)$ of $f$.
To reconstruct $f$ from $(C_0,\dots,C_{\ell-1})$, we need in addition the
following information.
\begin{defn}
The \emph{auxiliary sequence} for an ordered orbit decomposition $(C_0,\dots,C_{\ell-1})$
of a function $f$ is $(p_0,\dots,p_{\ell-1})$, where for each $i=0,\dots,\ell-1$,
$p_i$ is the position of $f(v_i)$ in the concatenated sequence $C_0C_1\dots C_{\ell-1}$,
$v_i$ being the last element in the sequence $C_i$.
\end{defn}

\begin{exam}\label{ex1}
Consider the function $f$ on $\{0,\dots,6\}$ whose graph is
$$\XYmatrix{
0 \ar[r] & 5 \ar@/^1pc/[rr] &               & 2 \ar@/^/[ld] & 4\ar[l] & 6\ar@/^/[d]\\
         &                  & 3 \ar@/^/[lu] &                &        & 1 \ar@/^/[u]
}$$
The ordered orbit decomposition of $f$ is
$$(C_0,C_1,C_2)=((0,5,2,3),(1,6),(4)),$$
and the ordered orbit structure is $(|C_0|,|C_1|,|C_2|)=(4,2,1)$.
$C_0$ and $C_1$ are $\rho$-orbits, whereas $C_2$ is not.
The concatenated orbits $C_0C_1C_2$ give $(0,5,2,3,1,6,4)$.
Now, $3$ is the last element in $C_0$, and the position of $f(3)=5$ in
the concatenated sequence is $1$. $6$ is the last element in $C_1$, and the position of $f(6)=1$ in
the concatenated sequence is $4$. Similarly, the position of $f(4)=2$ is $2$, so
the auxiliary sequence is $(1,4,2)$.
\end{exam}

\begin{defn}\label{fffunc}
Assume that $(m_0,m_1,\dots,m_{\ell-1})$ is
the ordered orbit structure of a function $f$ on $\Nset$,
and that the auxiliary sequence is $(p_0,\dots,p_{\ell-1})$.
For each $i = 0,\dots,\ell-1$, let $s_i = m_0+\dots+m_i$.
The \emph{fast forward function coded by $(m_0,m_1,\dots,m_{\ell-1})$ and $(p_0,\dots,p_{\ell-1})$}
is the function $\pi:\Nset\to\Nset$
whose ordered orbit decomposition is
$$((\ub{0\dots s_0-1}_{m_0}),(\ub{s_0\dots s_1-1}_{m_1}),(\ub{s_1\dots s_2-1}_{m_2}),\dots,
(\ub{s_{\ell-2}\dots N-1}_{m_{\ell-1}})),$$
and whose auxiliary sequence is $(p_0,\dots,p_{\ell-1})$.
\end{defn}

\begin{exam}\label{ex2}
The ordered orbit structure of $f$ in Example \ref{ex1} is
$(4,2,1)$, and the auxiliary sequence is $(1,4,2)$.
The fast forward function $\pi$ corresponding to $f$ is that with
the same auxiliary sequence and whose ordered orbit decomposition
is $((0,1,2,3),(4,5),(6))$. The graph of $\pi$ is
$$\XYmatrix{
0 \ar[r] & 1 \ar@/^1pc/[rr] &               & 2 \ar@/^/[ld] & 6\ar[l] & 4\ar@/^/[d]\\
         &                  & 3 \ar@/^/[lu] &                &        & 5 \ar@/^/[u]
}$$
Using the auxiliary sequence we have, e.g., that
$$\pi^{10}(6)=\pi^9(2)=\pi^7(1)=1+(7 \bmod 3) = 2,$$
as can be verified directly.
\end{exam}

Example \ref{ex2} hints to the following recursive procedure to compute $\pi^m(x)$.
Again, let $i(x)$ be such that $s_{i(x)}\le x<s_{i(x)+1}$
for each $x=0,\dots,N-1$.
\be
\itm Let $r = m-(s_{i(x)+1}-x)$. (Note that $r<m$.)
\itm If $r<0$, then $\pi^m(x) = x+m$.
\itm Else:
\be
\itm If $s_{i(x)}\le p_{i(x)}$ then $C_{i(x)}$ is a $\rho$-orbit,
and therefore
$$\pi^m(x)=p_{i(x)} + (r \bmod (s_{i(x)+1}-p_{i(x)})).$$
\itm Otherwise, $\pi^m(x)=\pi^{r}(p_{i(x)})$.
\ee
\ee
Case (b) is the only case where a recursion is made. Note that in this case, $p_{i(x)}<s_{i(x)}$,
i.e.\ we descend to a previous component.
We therefore call this case a \emph{descent}.

For simplicity, use the term \emph{basic operation}
for either a basic arithmetic operation, a comparison, or a lookup access.
It follows that each descent requires less than $10$ basic operations.

\begin{cor}\label{des}
The complexity of evaluating $\pi^m(x)$ is a constant $c\le 10$ times the
number of descents needed until a $\rho$-orbit is reached.
\end{cor}

\begin{rem}
In the sequel, we will measure the complexity by the number of descents.
The constant $c$ by which this should be multiplied (Corollary \ref{des})
can be made smaller by pre-computing lookup tables for $s_{i(x)}$, $p_{i(x)}$,
and $s_{i(x)+1}-p_{i(x)}$.
\end{rem}

We now describe the basic method for coding $f$ as a fast forward function.
The running time of this transformation is a small constant multiple
of $N$.

\begin{code}\label{fcode}
Assume that $f$ is a function on $\Nset$. Code $f$ as follows.
\be
\itm Compute the ordered orbit decomposition of $f$:
$$((\ub{b_0\dots b_{s_0-1}}_{m_0}),(\ub{b_{s_0}\dots b_{s_1-1}}_{m_1}),(\ub{b_{s_1}\dots b_{s_2-1}}_{m_2}),\dots,
(\ub{b_{s_{\ell-2}}\dots b_{N-1}}_{m_{\ell-1}})).$$
\itm Define a permutation $\sigma$ on $\Nset$ by $\sigma(x) = b_x$ for each $x=0,\dots,N-1$.
\itm Use $\sigma\inv$ to compute the auxiliary sequence $(p_0,\dots,p_{\ell-1})$.
\itm Store in memory the following tables: $\sigma$, $\sigma\inv$, the list $s_0,\dots,s_{\ell-1}$
(where $s_k=m_0+\dots+m_k$ for each $k$), the auxiliary sequence $(p_0,\dots,p_{\ell-1})$,
and the assignment $x\mapsto i(x)$ (such that each $x\in C_{i(x)}$).
\ee
\end{code}
Note that the code of $f$ defines the fast forward function $\pi$ coded by $(m_0,\dots,m_{\ell-1})$
and $(p_0,\dots,p_{\ell-1})$, and that $f=\sigma\circ\pi\circ\sigma\inv$. Thus,
$$f^m(x)=\sigma(\pi^m(\sigma\inv(x))$$
for each $x$ and $m$.
Consequently, if the maximal number of descents in $\pi$ is small, $f^m(x)$ can be evaluated
efficiently for all $m$ and $x$.

Simulations show that for random functions $f$, the maximal number of descents in
the evaluations $f^m(x)$ is around $\log_2 N$. We will give concrete results
for a better approach in the sequel.

\subsection{An improved approach}\label{SecFuncImp}

There are pathological cases where the number of descents can be $N$.
We exhibit the extreme case, with a hint concerning how it can be
avoided.

\begin{exam}\label{ex3}
Consider the function $f(k)=\max\{0,k-1\}$:
$$\XYmatrix{
0\ar@(ul,dl) & 1 \ar[l] & 2 \ar[l] & \dots \ar[l] & (N-2) \ar[l] & (N-1)\ar[l]
}$$
The ordered orbit decomposition of $f$ is $((0),(1),(2),\dots,(N-1))$,
and the auxiliary sequence is $(0,0,1,2,\dots,N-2)$.
The ordered orbit structure is $(1,1,\dots,1)$, and
the corresponding fast forward function $\pi$ is equal to $f$.
Computing $\pi^m(N-1)$ for $m\ge N-1$ requires $N-1$ descents.

Now consider the function $g(k)=\min\{k+1, N-1\}$:
$$\XYmatrix{
*{(N-1)}\ar@<3ex>@(ul,dl) & (N-2) \ar[l] & \dots \ar[l] & 2 \ar[l] & 1 \ar[l] & 0\ar[l]
}$$
The ordered orbit decomposition of $g$ is $((0,1,2,\dots,N-1))$,
and the auxiliary sequence is $(0)$.
The ordered orbit structure is $(N)$, and
the corresponding fast forward function $\pi$ is equal to $g$.
No descents at all are required to compute values $\pi^m(x)$.
\end{exam}

The following definition captures the improvement made in the
second part of the last example.

\begin{defn}\label{GCSf}
Assume that $f$ is a function on $\Nset$.
The \emph{greedy orbit decomposition} of $f$ is the sequence
$(C_0,\dots,C_{\ell-1})$ defined as follows, where
a \emph{maximal} orbit is an orbit of maximal length,
and when there is more than one maximal orbit, we choose the one
starting with the least point:
\be
\itm $C_0$ is the maximal orbit in $G$.
\itm For $k>0$, if $V\neq C_0\cup C_1\cup \dots \cup C_{k-1}$,
then $C_k$ is the maximal orbit in the subgraph
induced by $G$ on the vertices in $V\sm (C_0\cup\dots \cup C_{k-1})$.
\itm $\ell$ is the least $k$ such that $V=C_0\cup\dots \cup C_{k-1}$.
\ee
The \emph{greedy orbit structure} of $f$ is the sequence $(|C_0|,\dots,|C_{\ell-1}|)$.
\end{defn}

\begin{rem}
Given a graph of a function on $\Nset$,
one can attach to each vertex the length of its orbit.
This can be done in $\le 2N$ steps.
After removing an orbit from the graph, only the points
which eventually enter the orbit need to be modified.
Even if we recompute all lengths after each removal of an
orbit, the overall complexity is not more (and usually much less) than
$$2N+2(N-1)+\dots+2\approx N^2.$$
Since the procedure is done only once and offline,
we do not try to optimize further.
\end{rem}

Having defined the greedy orbit decomposition of $f$, we can proceed to define, with respect to it,
the auxiliary sequence and the other definitions, as well as the coding, exactly as in
Section \ref{SecFuncBas}.

\begin{exam}\label{ex4}
Notation as in Example \ref{ex3}, we have that
the greedy orbit decomposition of $f$ is $((N-1,N-2,\dots,1,0))$,
the auxiliary sequence is $(N-1)$, and the ordered orbit structure is $(N)$.
The fast forward function $\pi$ is equal to $g$, and
no descents at all are required to compute values $\pi^m(x)$.
\end{exam}

The following theorem shows that,
using the greedy orbit structure, the maximal possible number of
descents cannot be greater than about $\sqrt{2N}$.

\begin{thm}\label{det}
Assume that $f$ is a function on $\Nset$.
Then the maximal number of descents in the greedy orbit structure of $f$
is not greater than $\lfloor(\sqrt{1+8N}-3)/2\rfloor$.
\end{thm}
\begin{proof}
Consider the greedy orbit structure $(C_0,\dots,C_{\ell-1})$ and auxiliary sequence
$(p_0,\dots,p_{\ell-1})$ for $f$.
Let $d$ be the maximal number of descents in this structure.
Then there is a sequence $i_0<i_1<\dots<i_d$ such that for each $j=1,\dots,d$,
the last member in $C_{i_j}$ is mapped by $f$ to some member of $C_{i_{j-1}}$.
Since $(C_0,\dots,C_{\ell-1})$ is a greedy orbit structure, we have
that
$$|C_{i_0}|>|C_{i_1}|>\dots>|C_{i_d}|.$$
Indeed, for each $j=1,\dots,d$,
as $C_{i_j}$ is not a $\rho$-orbit,
the orbit in $\la V\sm (C_0\cup\dots \cup C_{i_{j-1}-1}),E\ra$
starting with the first element of $C_{i_j}$
is of size at least $|C_{i_j}|+1$, and by the maximality
of $|C_{i_{j-1}}|$, we have that $|C_{i_j}|+1\le|C_{i_{j-1}}|$.

Consequently, for each $j=0,\dots,d$, $|C_{i_j}|\ge d-j+1$,
and therefore
$$N=|V|\ge\left|\bigcup_{j=0}^d C_{i_j}\right|=\sum_{j=0}^d |C_{i_j}|\ge \sum_{j=0}^d (j+1) = \frac{(d+1)(d+2)}{2}.$$
Thus, $d^2+3d+(2-2N)\le 0$, that is,
$$d\le\frac{-3+\sqrt{9-4(2-2N)}}{2} = \frac{\sqrt{1+8N}-3}{2}.\qedhere$$
\end{proof}

The bound in Theorem \ref{det} cannot be improved.

\begin{exam}\label{ex5}
Fix $N$. Let $d=\lfloor(\sqrt{1+8N}-3)/2\rfloor$.
Then $M=(d+1)(d+2)/2\le N$. We will define a function
on $\{0,\dots,M-1\}$ whose greedy orbit decomposition has
$d$ descents starting at $M-1$.
Clearly, such a function can be extended to a function on
$\Nset$ with $d$ descents in its greedy orbit decomposition
by extending the first component.

Consider the function $f$ whose greedy orbit decomposition is
$$((0,1,\dots,d),\dots,(M-6,M-5,M-4),(M-3,M-2),(M-1))$$
with auxiliary sequence $(d,d,\dots,M-4,M-2)$.
There are $d+1$ components, and each component
is descended into the previous component, so starting
at the value $M-1$ we have $d$ many descents.

E.g., for $d=3$, $M=10$ and the function is
$$\XYmatrix{
0 \ar[dr] & & 4 \ar[dr] & & 7 \ar[dr] & & 9 \ar[dl] \\
& 1 \ar[dr] & &  5 \ar[dr] & & 8 \ar[dl]\\
& & 2 \ar[dr] & & 6 \ar[dl]\\
& & & 3
}$$
its greedy orbit decomposition is
$((0,1,2,3),(4,5,6),(7,8),(9))$ and the
auxiliary sequence is $(3,3,6,8)$.
Computing $f^{3}(9)$ requires $3$ descents.
\end{exam}

Note that Example \ref{ex5} has an orbit decomposition with
at most one descent. E.g., in the case $d=3$ we can take
$((9,8,6,3),(0,1,2),(4,5),(7))$ with auxiliary sequence
$(3,3,2,1)$.

This suggests that in Definition \ref{GCSf}, when we have more than one maximal
orbit, we should try all possibilities.
This way, the algorithm becomes exponential.
We have tried a randomized approach which broke ties using coin flips.
It did not give significantly better results.
We would be glad but surprised if the answer to
the following would turn out positive.

\begin{prob}\label{prob1}
Does there exist an efficient algorithm to find, for a given function $f$ on $\Nset$,
an orbit decomposition for which the maximal number of descents is
as small as it can be for $f$?
\end{prob}

\subsection{The random case}
The random case, and presumably most of the cases encountered in practice, behaves
much better than is provable for the worst case.
For each $N=2^2,2^3,\dots,2^{20}$, we have sampled $100$ random functions on $\Nset$.
For these, we have computed the maximum and average number of descents.
The results appear in Figure \ref{numdec}.
\begin{figure}[!h]
\begin{changemargin}{-2.5cm}{-2cm}
\begin{center}
\Includegraphics{width=13cm, height=10cm}{numdec4}
\end{center}
\end{changemargin}
\caption{Number of descents in the random case.}\label{numdec}
\end{figure}
Figure \ref{numdec} contains three increasing and one decreasing graphs.
Among the increasing graphs,
the uppermost is just $\log_2N$, the intermediate graph is the maximum
number of descents encountered for each $N$,
and the lowest is the average number of descents.
The decreasing graph is $\log_2N$ divided by the average
number of descents.

An interesting observation is that none of the samples contained a point
with more than $\log_2N$ many descents. This should be contrasted with
Example \ref{ex5}, and suggests that the cases in which the complexity
of evaluating $f^m(x)$ can be larger than about $\log_2N$ are indeed pathological.

Another observation, which is of great practical interest, is
supplied by the decreasing graph: It shows that for the checked
values of $N$, and presumably for all practical values of $N$, the
average number of descents is about $(\log_2N)/5$ or less.
Recall from Corollary \ref{des} and the remark after it,
that the overall complexity is a small multiple of this number.

\medskip

We conclude the paper by demonstrating that the mere consideration of
average complexity rather than maximal complexity
does not suffice to obtain the logarithmic phenomenon which we encountered
in the random case.

\begin{exam}
Consider the function $f$ described in Example \ref{ex5},
and assume that $(\sqrt{1+8N}-3)/2$ is an integer
(otherwise the following is only approximate).
In this case, $d$ is equal to this number,
and $(d+1)(d+2)=2N$.

The average number of descents for $f$ is $1/N$ times
\begin{eqnarray*}
\lefteqn{1\cdot d+2\cdot(d-1)+\dots+d\cdot 1=}\\
& = & \sum_{i=1}^d i(d+1-i) = \sum_{i=1}^d i(d+1) - \sum_{i=1}^d i^2 =\\
& = & \frac{(d+1)(1+d)d}{2} - \frac{d(d+1)(2d+1)}{6} =\\
& = & \frac{d(d+1)(d+2)}{6}=N\cdot\frac{d}{3}.
\end{eqnarray*}
Thus, the average number of descents in $f$ is
$d/3$ (which is roughly $\sqrt{2N}/3$).
\end{exam}

\section{Conclusions, improvements, and open problems}

We have shown that every lookup table $T$ of size $N$
can be coded by $cN$ elements where $c$ is a small constant,
such that computations of the form $T^m(x)$---the $m$th iterate of
$T$ at $x$---can be done efficiently.
The efficiency is measured in the number of recursions
(descents) which our algorithm performs.

In the case that $T$ is a permutation, no recursions are needed.
When $T$ is a general function, we can have up to
$\sqrt{2N}$ recursions but not more than that,
and if $T$ is random, then the number of recursions reduces
to about $(\log_2N)/5$. The last assertion was only verified
experimentally, and a rigorous explanation of this
reduction from $O(\sqrt{N})$ to $O(\log N)$ in the
random case would be interesting.

In a work in progress with Yossi Oren, we introduce another
heuristic for the decomposition of graphs of functions.
For this heuristic, the maximal number of descents
reduces to $\log_2 N$ (which is optimal with respect to the worst-case behavior).
There are still cases where the heuristic described in the current paper
outperforms the newer heuristic, though.\footnote{\label{newfootnote}This paragraph appears in
the published version of this paper, but was not noticed by
Tsung-Hsi Tsai, who in his paper \emph{Efficient computation of the iteration of functions},
to appear in Theoretical Computer Science, was able to reconstruct our new heuristic and
prove that it is optimal. In particular, we were mistaken to think that there
are cases where the new heuristic is not optimal.}

The task of finding a heuristic approach which reduces the
average number of recursions in the computations
$T^m(x)$ to less than our $(\log_2 N)/5$ seems to be of great
practical interest.

\ed